# Low-noise supercontinuum generation in chiral all-normal dispersion photonic crystal fibers

**MARKUS LIPPL,**[1,2*] **MICHAEL H. FROSZ,**[1] **NICOLAS Y. JOLY**[2,1,3]

[1]*Max Planck Institute for the Science of Light, Staudtstraße 2, 91058 Erlangen, Germany*
[2]*Department of Physics, Friedrich-Alexander-Universität, Staudtstraße 2, 91058 Erlangen, Germany*
[3]*Interdisciplinary Centre for Nanostructured Films, Cauerstr.3, 91058 Erlangen, Germany*
*\*Corresponding author: markus.lippl@mpl.mpg.de*

*Abstract* We present the advantages of supercontinuum generation in chiral, therefore circularly birefringent, all-normal dispersion fibers. Due to the absence of nonlinear power transfer between the polarization eigenstates of the fiber, chiral all-normal dispersion fibers do not exhibit any polarization instabilities and thus are an ideal platform for low-noise supercontinuum generation. By pumping a chiral all-normal dispersion fiber at 802 nm, we obtained an octave-spanning, robustly circularly polarized supercontinuum with low noise.

Solid-core photonic crystal fibers (PCFs), which offer enhanced control of the nonlinearity as well as an extensively engineerable dispersion landscape [1], have rapidly become the workhorse for supercontinuum (SC) generation since its first demonstration in PCF in 1999 [2]. Because the spectra of these ultra-broadband sources can spread over multiple octaves while maintaining very good spatial coherence, they are an ideal tool for various applications ranging from optical coherence tomography [3] to high-resolution spectroscopy [4,5]. Although the capability of significantly broadening the spectrum of an initially narrowband source is appealing for many applications, the noise properties of such sources can be particularly detrimental for others and many groups worldwide currently focus on improving the noise characteristics of such sources [6]. Ultra-broad supercontinua based on soliton fission dynamics accompanied by the emission of dispersive waves are known to be very noisy [7]. In order to maintain high temporal coherence, one needs to operate in the normal dispersion regime so as to avoid the emergence of solitons, or other noise-seeded processes such as Raman scattering or modulational instability [7]. In all-normal dispersion (ANDi) fibers the broadening mechanism relies on self-phase modulation (SPM) and optical wave breaking (OWB) [8], which are both coherent effects. The impact of these effects can be maximized by tailoring the dispersion to be small and flat in order to minimize the effect of temporal broadening and maximize the impact of the nonlinearity. Nevertheless, there remain some sources of incoherence and noise such as polarization instabilities [9] and polarization modulation instability [10]. These issues are usually addressed by using fibers exhibiting high linear birefringence [11] provided that the pump is polarized along the slow axis of the fiber but not along the fast axis. This is due to the coherent coupling between the two linearly polarized eigenstates which makes the fast axis unstable [9]. Here, we use a totally different approach by utilizing a chiral solid-core PCF, which supports robustly circularly polarized modes. We recently showed that such a PCF could produce an almost perfectly circularly polarized supercontinuum when pumped with circularly polarized light [12]. We show here that by utilizing a chiral solid-core ANDi PCF, not only can we benefit from the highly coherent SPM-driven broadening processes, but we can also fully eliminate the possible coupling between the slow and fast axes present in commonly used birefringent fibers.

Solid-core chiral PCFs consist of a hexagonal lattice of holes that evolves helically around the center. Such a fiber can be fabricated by twisting the preform during the fiber drawing process. The consequence of the resulting chirality is that the fiber can exhibit strong optical activity, so that the fiber becomes circularly birefringent [13]. In this case, the generalized nonlinear Schrödinger equation (GNLSE) expressed in the co-moving time frame $T$,

$$\frac{\partial A_p(z,T)}{\partial z} = \mathcal{D}^{(p)}(z,T) + \mathcal{N}^{(p)}(z,T), \quad (1)$$

which describes the spatio-temporal propagation of the amplitude $A_p$ of the mode $p$ over a distance $z$, includes a nonlinear operator that reads

$$\mathcal{N}^{(l,r)}(z,T) = i\frac{2}{3}\gamma\left(1 + i\tau\partial_T\right) \times \left(|A_{l,r}|^2 + 2|A_{r,l}|^2\right)A_{l,r} \quad (2)$$

where the Raman-scattering term is neglected [14,15]. The linear operator $\mathcal{D}^{(p)}(z,T)$ in the GNLSE describes both the loss and the dispersion. The indices $l$ and $r$ indicate the left- and right-hand circularly polarized (LHCP and RHCP) eigenstates [15]. The nonlinear parameter is given by

$$\gamma = \frac{\omega_0 n_2 n_{eff}(\omega_0)}{c n_{eff}(\omega)\sqrt{A_{eff}(\omega)A_{eff}(\omega_0)}}, \quad (3)$$

with $\omega_0$ the central angular frequency, $n_{eff}$ the effective index of the mode, $c$ the speed of light in vacuum and $A_{eff}$ the effective area of the mode, and $n_2$ is the nonlinear refractive index of silica [10]. $\tau=1/\omega_0$ is the shock time constant. The terms $|A_{l,r}|^2 A_{l,r}$ and $|A_{r,l}|^2 A_{l,r}$ describe respectively the self- (SPM) and cross-phase modulation (XPM) [15]. This nonlinear operator contrasts with the case when linear birefringence is considered

$$\mathcal{N}^{(h,v)}(z,t) = i\gamma(1 + i\tau\partial_t)$$
$$\times \left[\left(|A_{h,v}|^2 + \frac{2}{3}|A_{v,h}|^2\right)A_{h,v} + \frac{1}{3}A_{v,h}^* A_{h,v}^2\right]. \quad (4)$$

Here, $h$ and $v$ indicate the horizontal and vertical polarization. First of all, the relative strength of XPM is twice as high as the SPM contribution in circularly birefringent fibers, as opposed to 2/3 in the case of linear birefringence. Moreover, we have a reduction of the nonlinearity by a factor of 1/3. Most important is the additional term $A_{v,h}^* A_{h,v}^2$, which does not appear in the case of circularly birefringent fibers and describes coherent coupling between the orthogonal modes. This induces energy transfer between the linearly polarized eigenstates and leads to intermodal four-wave mixing, polarization instabilities, as well as polarization modulational instability [9,15]. It is the origin of the asymmetric behavior when either the fast or slow axis is used. Pumping along the fast axis can lead to polarization instabilities in contrast to the slow axis. Since the coherent coupling term is absent for circularly birefringent fibers, we can assume that such fibers are immune against such instabilities and are the ideal platform for low-noise supercontinuum generation.

To verify the effects described above, we designed and fabricated a chiral ANDi fiber using the stack-and-draw technique. The fiber is made of undoped silica. In order to obtain a flat dispersion, while being in the normal dispersion regime over the whole range of interest, a hole pitch $\Lambda$ of 1.3 μm and a hole diameter $d$ of 0.6 μm was chosen so that the $d/\Lambda$ is 0.46. This makes the fiber practically single mode over the wavelength range of interest [16]. To ensure good confinement, particularly in the infrared, we used nine rings of holes around the core so as to decrease waveguide losses. To make sure that the fiber exhibits pure circularly polarized eigenstates, it is necessary to keep the linear birefringence minimal. Therefore, the capillaries used for the stack were selected so as to have the diameter varying less than ±0.5%. The chosen twist period in the fiber is 5.3 mm.

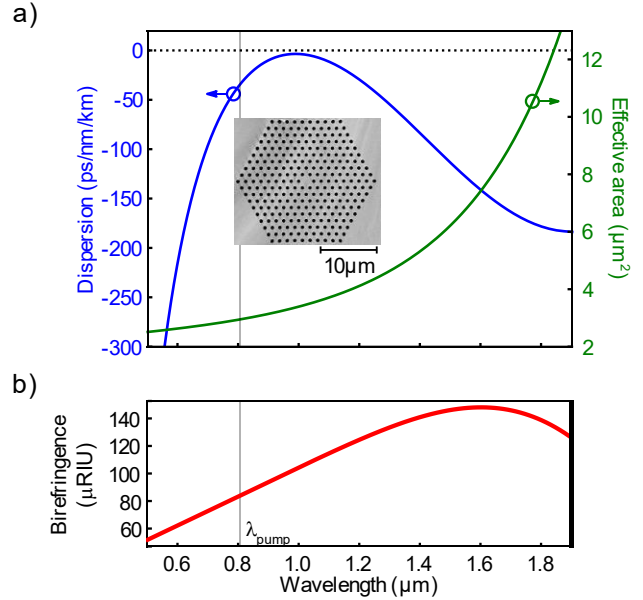

Fig. 1. Properties of the manufactured fiber obtained from finite element modelling. a) Dispersion (blue, left axis) and effective mode area (green, right axis) as a function of wavelength. The inset shows the scanning electron micrograph used for the modelling. The grey vertical line indicates the pump wavelength used in this work. b) Circular birefringence of the fiber as a function of wavelength.

Based on a high-resolution scanning electron micrograph (see inset of Fig. 1) and the twist period, dispersion, effective mode area and birefringence were modelled using finite element calculation. As shown in Fig. 1, the dispersion is normal over the whole range considered. On the short-wavelength side of the pump the material dispersion is dominant, while on the long-wavelength side the waveguide provides the strongest contribution. The birefringence of the fiber is - for all practical purposes - purely circular and strong. The beat length is 9.6 mm at 800 nm wavelength, which is an order of magnitude below comparable linearly birefringent ANDi fibers (75 mm at 1500 nm wavelength [11]).

A schematic of the experimental setup is shown in Fig. 2. The pump is an amplified Ti:Sapphire laser operating at 802 nm with a pulse duration of 165 fs. The repetition rate is 250 kHz and the available average power is 100 mW. An autocorrelation trace is shown in the inset of Fig. 2. The power of the pump is adjusted using a half-wave retarder in conjunction with a polarizing beam splitter, which also has the function of providing a well-defined linear polarization state. We generate the specific polarization states through a combination of half- and quarter-wave plates placed prior to the in-coupling lens. Using a coated aspheric lens, the pump is launched into a 150 mm long piece of fiber, which is kept straight. The output of the fiber is collimated with another aspheric lens. A pair of flip mirrors allows steering the beam towards different measurement devices. The generated supercontinuum can be sent directly to a photodiode (Alphalas UPD-40-UVIR-P, flip mirror i) or an optical spectrum analyzer (OSA) using a multimode fiber (flip mirror ii). Alternatively, it can pass through a half-wave plate and a Glan-Taylor polarizer (GP) before it is sent to the OSA.

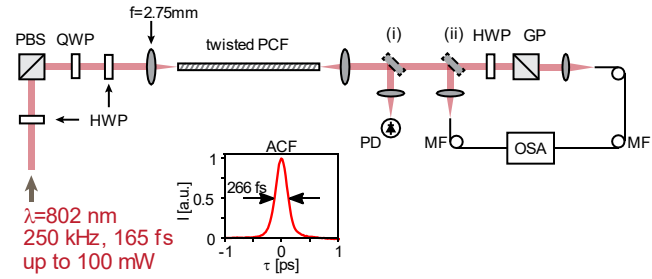

Fig. 2. Schematic of the setup used for the measurements presented in this work. HWP: Half-wave plate, PBS: polarizing beam splitter, QWP: quarter-wave plate; GP: Glan-Taylor polarizer, PD: photodetector, MF: multimode fiber, OSA: optical spectrum analyzer. (i) and (ii) denote flip mirrors which can be set optionally. The autocorrelation function (ACF) of the pump source is shown in the inset.

The measured energy-dependent evolution of the spectrum is shown in Fig. 3 for a left-hand circularly polarized pump. The initial broadening is governed by SPM. Almost from the beginning there is a distinct spectral band (~ 650 nm) at the short-wavelength side which is due to self-steepening and resulting shock formation [8].

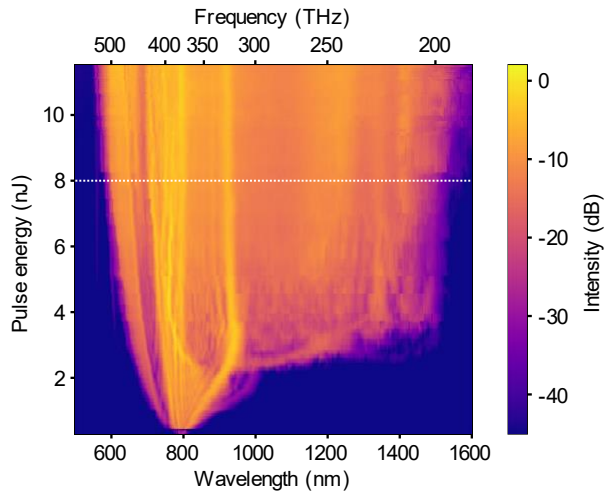

Fig. 3. Power dependent measured spectral output for LHCP pumping showing the SPM-like broadening at low pulse energies. The extension to the infrared is resonant radiation. The distinct band at the short-wavelength side is due to shock formation. The dashed white line indicates an input energy of 8 nJ, for which the spectrum and polarization state is presented in Fig. 4.

The asymmetry of the SPM-ripples leading to a peak at the long-wavelength side is due to the pump wavelength being located on the slope of the dispersion curve. At around 2.4 nJ, the spectrum extends up to ~1 μm, which corresponds to the top of the dispersion curve, where the dispersion almost vanishes (Fig. 1(a)). At this point, the system shows rapid generation of resonant radiation [17] extending up to 1540 nm. For a pulse energy of 8 nJ the spectrum and polarization properties are summarized in Fig. 4. The spectra of the pump laser itself as well as the supercontinua for both RHCP and LHCP pumping are shown in Fig. 4a. The supercontinua extend from 590 nm to 1500 nm covering more than 1.3 octaves.

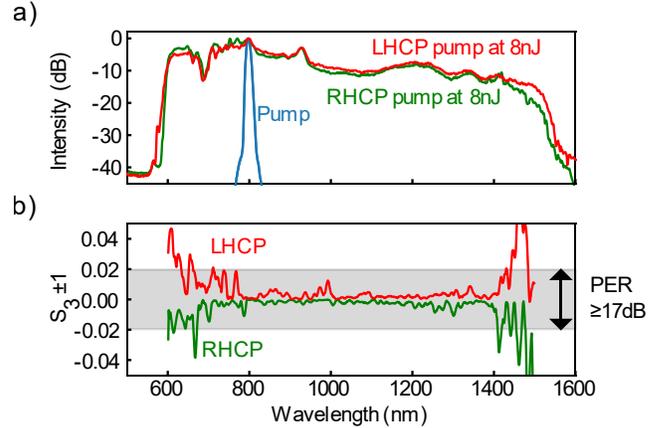

Fig. 4. (a) Measured spectra of the pump (blue) and the supercontinua when pumped with LHCP (red) or RHCP (green) pump obtained with a 150 mm long fiber at a pulse energy of 8 nJ. (b) Measured ellipticity (Stokes parameter $S_3$) of the supercontinuum for the same parameters (150 mm, 8 nJ) with the pump being LHCP (red) or RHCP (green). The grey area indicates a polarization extinction ratio of 17 dB or better.

To show that the polarization state of the generated supercontinuum is circular over the whole spectral range, we performed spectrally resolved polarimetry by projecting the polarization state of the supercontinuum onto respectively vertical, horizontal, diagonal and anti-diagonal polarization using a half-wave plate and a Glan-Taylor polarizer (GP). The spectra of these projected supercontinua are measured by the OSA. From this measurement, we obtain the spectral intensity of the supercontinuum (identical to $S_0$) as well as the first two components of the Stokes vector ($S_1$ and $S_2$). Since the underlying broadening mechanism of the supercontinuum is mainly SPM, it is coherent [7] and we can deduce the third Stokes component $S_3$. For RHCP (LHCP) pumping one would expect $S_3$ to be +1 (-1), respectively [12]. The deviation from this expectation is shown in Fig. 4b. One can see, that the supercontinuum is robustly circularly polarized over the whole spectrum, having a polarization extinction ratio of 17 dB or better. The polarization extinction ratio is independent of whether the pump is LHCP or RHCP, which corresponds to a polarization eigenstate of the fiber. This stands in contrast with the case of supercontinua generated in linearly birefringent fiber where using the fast axis does not provide a robust polarization state [9]. At the edges of the supercontinuum, the polarization deviates more. This can be attributed to a worse signal-to-noise ratio as well as a drift of coupling efficiency between the measurements for the individual projections (on a timescale of minutes).

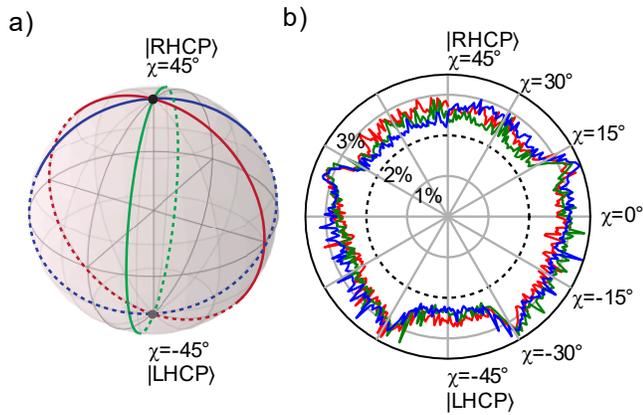

Fig. 5. Relative intensity noise (RIN) measurement. a) Poincaré sphere with polarization eigenstates of the fiber highlighted (RHCP and LHCP). The meridians are traces of polarization states for which the relative intensity noise is measured. b) Relative intensity noise depending on pump ellipticity $\chi$ for different azimuths separated by 36°. The RIN is shown in the radial direction. The angles are the inclination angles of the pump state on the Poincaré sphere. The dashed line is the RIN of the pump laser of 1.97%.

To characterize the influence of the pump polarization on the noise characteristics, namely the relative intensity noise (RIN), the noise was measured for different pump polarizations from RHCP to LHCP along three different meridians on the Poincaré sphere (Fig. 5a) [11]. The difference in azimuth of the chosen meridians is 36° and was chosen to not coincide with the six-fold rotational symmetry of the fiber. We measure the noise properties by detecting the fluctuations of total intensity using a photodiode and an oscilloscope, as described in [18] (flip mirror i) along each meridian. For every RIN measurement, 8000 pulses were recorded with the oscilloscope. The RIN is then the ratio between the standard deviation and the mean of the pulse amplitudes measured on the oscilloscope [18]. As one can see, the RIN coincides for each meridian; it does not depend on the azimuth of the pump polarization. Both the right- and the left-handed eigenstates are equal in terms of noise. This clearly indicates that both the fast and the slow axis of the chiral fiber are equally suitable for low-noise supercontinuum generation. The absence of any strong, narrow noise peaks, which were previously reported in the case of linearly birefringent fibers, shows that there are no additional noise mechanisms such as cross-phase modulation instability or polarization modulation instability present [11]. The region between +15° and -30° of chirality shows a higher amount of noise, which is due to slight walk-off of the pump beam and residual pointing fluctuations of the used laser. The generally high noise level is due to the comparatively high RIN of the pump laser (1.97%) as well as residual pointing instabilities left after beam stabilization.

In conclusion, we reported on supercontinuum generation in a chiral all-normal dispersion fiber. The measurements of the noise performance of the system confirm that for circularly birefringent fibers, the orthogonally polarized LHCP and RHCP eigenstates exhibit a similar behavior and do not present any evidence of polarization-induced instability. This strongly contrasts with supercontinuum generated in linearly birefringent fibers. We expect that noise engineering as well as polarization state engineering of supercontinuum sources is possible and can be used to satisfy the growing need for low-noise supercontinuum sources.

**Acknowledgements.** M. L. is part of the Max Planck School of Photonics supported by BMBF, Max Planck Society, and Fraunhofer Society. We acknowledge the financial support by Deutsche Forschungsgemeinschaft (DFG) (Grants No. JO-1090/6-1).
**Disclosures.** The authors declare no conflict of interests
**Data availability.** Data underlying the results presented in this paper are not publicly available at this time but may be obtained from the authors upon reasonable request.